
\def\d#1/d#2{ {\partial #1\over\partial #2} }

\newcount\sectno

\newcount\subsectno

\def\subsect{\global\advance\subsectno by1 \the\sectno.\the\subsectno }
 \def\sect{\subsectno=0 \global\advance\sectno by1 \the\sectno }

\def\pdr{\partial}

\def\al{\alpha}
\def\be{\beta}
\def\ga{\gamma}
\def\de{\delta}
\def\eps{\epsilon}
\def\half{{1\over 2}}
\def\tr{\hbox{tr}}

\def\linebreak{\hfil\break}


\newcount\eqnumber
\def\beq{ \global\advance\eqnumber by 1 $$ }
\def\eeq{ \eqno(\the\eqnumber)$$ }
\def\n{\global\advance \eqnumber by 1\eqno(\the\eqnumber)}
\def\puteqno{
\global\advance \eqnumber by 1 (\the\eqnumber)}
\def\beqs{$$\eqalign}
\def\eeqs{$$}


\def\ifundefined#1{\expandafter\ifx\csname
#1\endcsname\relax}

\newcount\refno \refno=0  
\def\[#1]{
\ifundefined{#1}
\advance\refno by 1
\expandafter\edef\csname #1\endcsname{\the\refno}
\fi[\csname #1\endcsname]}
\def\refis#1{\noindent\csname #1\endcsname. }

\def\label#1{
\ifundefined{#1}
\expandafter\edef\csname #1\endcsname{\the\eqnumber}
\else\message{label #1 already in use}
\fi{}}
\def\(#1){(\csname #1\endcsname)}
\def\eqn#1{(\csname #1\endcsname)}

\baselineskip=15pt
\parskip=10pt



\magnification=1200
\def\un#1{\underline{#1}}
\def\sgn{\;{\rm sgn}\;}
\def\for{\;\;{\rm for}\;\;}

\centerline{\bf Solitons in a Bilocal Field Theory}

\centerline{{\sl R.J. Henderson} and {\sl S. G. Rajeev}}

 \centerline{Department of Physics and Astronomy,}
\centerline{ University of
Rochester,}
\centerline{Rochester,NY 14627}

\centerline{ABSTRACT}

We obtain a bilocal classical field theory as the large $N$ limit of the
chiral
 Gross--Neveu (or non--abelian Thirring) model. Exact classical solutions
 that describe topological solitons are obtained. It is shown that their mass
 spectrum agrees with the large $N$ limit of the  spectrum of the chiral
 Gross--Neveu model.
\vfill\break

We will study the large $N$ limit of the two--
dimensional fermionic system defined by the Lagrangian
\beq
     L=\bar q^i[-i\ga_\mu \pdr^\mu]q_i +{g^2\over 2N}\bar
q^i\ga_\mu q_j \bar q^j\ga^\mu q_i.
\eeq
Here $i=1,\cdots N$.
This model is a non--abelian version of the Thirring model.
 It is also known as the chiral Gross--Neveu \[gn]
model  since the interaction can be rewritten in the
following form using a Fierz identity.
\beq
     L=\bar q^i[-i\ga_\mu \pdr^\mu]q_i -
{g^2\over 2N}[\bar q^i q_i \bar q^j q_j-\bar q^i\ga_5 q_i \bar
q^j\ga_5 q_j].
\eeq
The precise definition of the model requires a
renormalization of the coupling constant.
  This theory  is asymptotically free, the dimensionless coupling constant
$g^2$ is replaced by a dimensional constant as the true parameter
upon renormalization.
This model has been solved exactly\[wiegmann],\[andrei] by
Bethe ansatz methods; i.e.,
its spectrum and the $S$-matrix are known explicitly.
 In the limit as $N\to
\infty$, it  should tend to an exactly integrable classical
theory. This classical theory will require a renormalization
in order to be well--defined ( the beta--function does not
vanish in the large $N$ limit). Still, we should be able to
understand the large $N$ limit of the mass spectrum  in
terms of the classical solutions. Such a  direct understanding
of the large $N$ limit of this type of theories has not been achieved yet, as
 pointed
out in  \[polyakov]. Berezin\[berezin] has given a bi--local formulation of
the large $N$ limit of our type of model; however, the mass spectrum was not
obtained in that reference. What amounts to the linear approximation to our
theory has been studied in  a recent paper \[ohta].

In this paper we will obtain the classical theory corresponding to the
large $N$ limit of this
model as well as obtain its mass spectrum, in agreement with
the known exact solutions. We will apply the
bilocal bosonization method\[twodhadron],\[sphqcd],\[others] as well as
the non--perturbative renormalization methods \[renqm]
developed in previous papers to this model. Having
established the validity of these methods in this exactly
solvable  context, these methods can in the future be used
in  more realistic models such as spherical QCD\[sphqcd].
In fact, one of the lessons of this paper is how to define
classical field theories which require a renormalization.
We believe this is the first example of  exact soliton solutions in a bilocal
 field theory. We also discover some new phenomena such as a topological
charge with a continuous range of values.

The large $N$ limit of our  model can be studied by the
technique of summing an appropriate class of Feynman
diagrams. As is clear from other examples,\[thooft] this
will produce a free field theory of small oscillations
around the vacuum, the analogues of mesons in  two dimensional QCD.
However, it is clear from those examples that the true large
$N$ limit is a highly non--linear classical theory. In
particular it can have classical solutions which are
topological solitons and hence are large deviations from the
 vacuum. The bi--local bosonization methods of Refs.
\[twodhadron] can construct the complete large $N$ limit and
describe even these topological solitons \[twodbaryon]. ( In
2dqcd they are the baryons.) Indeed we will see that in the
case of the chiral Gross--Neveu model, {\it all} the
particles in the spectrum are topological solitons. A
surprise is that the topologically conserved quantum number
 can take continuous values between $0$ and $2\pi$.
A completely precise mathematical description of  such a topological invariant,
with a continuous range,
requires  the theory of projectors  on von Neumann
algebras and is beyond the scope of this paper. The
technical apparatus necessary for this seems to exist
already \[araki].

Let us begin by understanding the symmetries of the above
lagrangian. First of all it has a global symmetry under
$U(N)\times U(N)$, the first (second)  $U(N)$ acting on the
left  (right)
components of $q$. This non--abelian part of this
chiral symmetry  is unbroken and the
particles of theory transform under the completely anti--
symmetric tensor representations of  $SU(N)\times SU(N)$. (
This is unlike in 2dqcd, where all particles are in the
trivial representation of $SU(N)$ due to confinement.)
If $r=1,\cdots,N-1$ is the rank of the tensor, the particles
transform under the representation $(r,r)$  of $SU(N)\times
SU(N)$
and have masses\[andrei],\[wiegmann]
\beq
     m_r=m_1{\sin{\pi r\over N}\over \sin{\pi\over N}}.
\eeq
The quantum number $r$ is also a conserved quantity of the
fermionic lagrangian, corresponding to the symmetry group $Z_N$.
We might understand this discrete symmetry by considering
the $U(1)$ sub--group of the global symmetry corresponding
to
\beq
     q_i\to e^{ i\alpha}q_i.
\eeq
If $\alpha$ is an integer multiple of ${2\pi  \over N}$, this
transformation is in fact a particular element in
$SU(N)\times SU(N)$. This $Z_N$ subgroup may viewed as
measuring the fractional part of fermion number, if $q_i$ is
assigned a fermion number of ${1\over N}$. ( This is
analogous to the `triality' of $SU(3)$ representations.)
The part of fermion number which is integral  can be viewed
as a separate conserved quantity, whose $U(1)$ symmetry group has
no overlap with $SU(N)\times SU(N)$.  We will thus decompose
fermion number into two conserved quantities, one taking
integer values on each completely occupied one--particle state
(which we will
call baryon number $B$ in analogy with 2dqcd) and another,
$\theta$ taking values $0,2\pi{1\over N},2\pi{2\over N},\cdots, 2\pi[1-{1\over
N}]$ corresponding to $Z_N$.

 In the Dirac sea, any one--particle  state that is completely
filled will
contain   $N$ fermions and will make no contribution to
the conserved charge $\theta$  of this $Z_N$ symmetry. If a state
is partially filled, its contribution will be equal to the
ratio of the number of occupied states to the number of
available states, which is an integer multiple of ${1\over
N}$. The conserved quantity ${r\over N}$ is precisely this
fractional part of the baryon number.
Obviously, the `fractional part'  of the fermion number can
add up to give an integer if we sum over many states. Less
obvious is that the integral part of the fermion number can
add up to a fractional value when summed over an infinite
number of states; this has to do with regularizations that
are necessary to make such a sum well--defined \[semniemi].

There are several possible ways of  taking the limit $N\to
\infty$. For example, we could take the limit keeping the
mass $m_1$ of the lightest particle fixed. Then, the
spectrum would consist of two infinite towers of particles
of equally spaced masses; one tower corresponding to
$r=1,\cdots, {N\over 2}$ and the other to $r=N-1,N-2, \cdots
{N\over 2}+1$. ( This is for $N$ even; there is a similar
expression for $N$
odd.)  A more interesting large $N$ limit will be
obtained if we take the limit keeping the mass $\mu$ of the
heaviest particle fixed. Then the mass differences between
particles tends to zero in the large $N$ limit and the
spectrum merges into a continuum.  (Such limits of the exact
solutions have been studied before  in a  somewhat different
context.\[fateevwieg].)
Also, in this limit, the variable $\theta$ takes a continuous
range of values between $0$ and $2\pi$
so that the mass spectrum becomes the continuous set,
\beq
     m_\theta=\mu\sin{\theta\over 2}, \;\;{\rm for}\;\;
0\leq\theta\leq 2\pi.
\eeq
We will try to understand this limit as a
classical field theory.  We will see that $\theta$ corresponds to
a topologically conserved quantity with a continuous range
of values between $0$ and $2\pi$  in the classical theory.

     First we will give a heuristic derivation of the
classical theory corresponding to the large $N$ limit of the
Thirring model. Then we will give a  more rigorous {\it ab
initio} definition of this classical theory and find some
static solutions to it. We regard the following discussion of
the large $N$ limit only as motivation for the study of this
classical theory. A more  rigorous justification would be in terms
of the quantization of the classical theory, along lines
discussed in \[twodhadron]. ${1\over N}$ will play the role of
$\hbar $ in this quantization.

     We will study the theory in hamiltonian form, in which
the field operators satisfy the equal time anti--commutation
relations
\beq
     [q^{\dag i}_\alpha(x), q_{j\beta}(y)]_+
=\delta_{\alpha\beta}\delta^i_j\delta(x-y).
\eeq
$\alpha,\beta=1,2$ are spin indices, which we will find
convenient to suppress usually.
If we ignore delicate issues of
renormalization, the hamiltonian can be expressed in terms of
the color singlet bilinear $M(x,y)=-{1\over N}:q_i(x)q^{\dag
i}(y):$ ( here, $M(x,y)$ is a $2\times 2$ matrix in spin space
and $:\;:$ denotes normal ordering with respect to the free
hamiltonian):
\beq
     H=\int \bigg[\tr (-i\ga_5)\big[{\pdr M(y,x)\over \pdr
y}\big]_{x=y}-
{g^2\over 2}\tr \big[M^2(x,x)-\ga_5 M(x,x)\ga_5
M(x,x)\big]\bigg]dx.
\eeq
Conventional large $N$ arguments can be used to show that the
 observables $M(x,y)$ will have quantum fluctuations of order
${1\over N}$, so that their time evolution will be described
by classical equations of motion. (One way to see this is to
note that their commutators are of order ${1\over
N}$\[twodhadron]. They form a representation of the infinite
dimensional unitary Lie algebra, also called $W_\infty$
algebra in the context of matrix models.) In a theory such as
2dqcd which has color confinement, these would be a complete
set of observables. However, the physical states of our theory
transform under non--trivial representations of color, so that
these are not a complete set of observables.

We can now see that $M(x,y)$ along with the charges of the
$SU(N)\times SU(N)$ symmetry are in fact a complete set of
observables. It is best to understand this fact in a
regularized context in which the position variables are
allowed to take only a finite number ( say $K$) of values.
The complete set of bilinears
$\Phi_{i\alpha}^{j\beta}(x,y)=-i[q^{i\alpha}(x),q^{\dag j\beta}
(y)]$ form the Lie algebra $\un{U}(2KN)$ under commutation.
The
 Fermionic Fock space carries an irreducible representation of
this Lie algebra. This Lie algebra contains
$\un{SU}(N)\times \un{SU(N)}\times \un{U}(K)$ as a subalgebra.
The generators of this subalgebra are the charges of the
global symmetry $\un{SU}(N)\times \un{SU}(N)$ and the
operators
$M(x,y)$ themselves. The key fact is that the representation
of $U(2KN)$ on the fermionic Fock space remains irreducible
with respect to this subalgebra. ( We omit a detailed proof in
the interest of brevity.)  Thus any operator that commutes
with the charges of the $SU(N)\times SU(N)$ symmetry as well
as $M(x,y)$ is a multiple of the identity: these together form
a complete set of observables. Thus the dynamics of the theory
reduces to a classical dynamics for $M(x,y)$; the generators
of the global symmetry never become classical, but they have
trivial dynamics, being conserved quantities. Thus we will for
the most part concentrate on the variables $M(x,y)$. These
arguments can probably be made rigorous in the case were $x,y$
have infinite range, but the techniques required probably
involve use of $C^*$ or von Neumann algebras. We
will not attempt it in this paper.


Any classical theory is described by a phase space ( manifold
of allowed initial conditions), a symplectic structure on this
manifold ( or, Poisson brackets of a complete set of dynamical
variables) and  a hamiltonian. In some cases, the phase space
is described as the set of solutions to some set of
constraints satisfied by some dynamical variable, rather than
directly in terms  of a co-ordinate system.

In our classical theory, the dynamical variables are self-
adjoint operators   $M$  on the complex Hilbert space
$L^2(R,C^2)$. ( This can be thought of the one-particle
Hilbert space of a Dirac fermion in 1+1 dimensional space-
-time.) $M$ can be described  in terms of its kernel,
\beq
     Mu(x)=\int M(x,y)u(y) dy
\eeq
where $M(x,y)$ is a $2\times 2$ matrix valued function ( in
general distribution) of the two variables $x,y$.  An
equivalent description is in terms of the `symbol', $\tilde
M(x,p)$, which is a Fourier transform of the kernel with
respect to the relative co--ordinate:
\beq
 \tilde M(x,p)=\int M(x+{y\over 2},x-{y\over 2})e^{-ipy} dy.
\eeq
Since $M$ is self--adjoint, $\tilde M(x,p)$ is real valued.
Spatial translation acts  as follows on these variables: $M(x,y)\to M(x+a,y+a)$
 and $\tilde M(x,p)\to \tilde M(x+a,p)$. We can regard $\tilde M(x,p)$ as an
infinite component classical field, labelled by a continuous internal index
$p$. Such bilocal field variables have been useful in formulating the large
$N_c$ limit of models for QCD.

In spite of the fact that
operators  on Hilbert spaces make an appearance, we emphasize
that ours
is still a classical theory, the dynamical variable of which
just happens to be an infinite dimensional `matrix' $M(x,y)$.
The Poisson brackets satisfied by these variables are,
\beqs{
     \{M_{\alpha}^{\beta}(x,y),M_{\ga}^{\de}(z,u)\}&=
\delta^{\be}_{\ga}\delta(y-z)[M_{\alpha}^{\de}(x,u)+
\eps_{0\al}^{\de}(x,u)]\cr
& -\delta^{\de}_{\al}\delta(u-x)[M_{\ga}^{\be}(z,y)+
\eps_{0\gamma}^{\beta}(z,y)].\cr
}\eeqs
Here $\alpha,\beta$ are spin indices which we usually
suppress.
Also,
\beq
     \eps_{0\al}^{\be}(x,y)=\ga_{5\al}^{\be}\int {dp\over
2\pi} e^{ip(x-y)}\sgn(p)
\eeq
is the kernel of an operator $\eps_0$ whose square is one.
It has the physical meaning of being sign of the massless Dirac operator;
the eigenspace of $\eps_0$ with eigenvalue 1 (or -1) is the space of  positive
(or non--positive ) energy one--particle states.

This algebra  is the central extension of the unitary Lie
algebra,
called the Lundberg--Kac--Petersen\[kacpet]
extension. This algebra has also acquired the name $W_\infty$
algebra in the context of matrix models. The above Poisson
brackets are just the commutation relations of the operators
$M(x,y)$ introduced previously, except that a factor of ${
i\over N}$  has been removed.
This is the appropriate prescription as ${1\over N}$ plays the
role of $\hbar$ in our classical limit. The central term
proportional to
$\eps_0$ appears because of the normal ordering of $M$ in the
quantum theory.

We will impose some conditions on the asymptotic
behaviour of $\tilde
M(x,p)$ as $|p|\to \infty$.  These follow from  the asymptotic
behaviour of the
matrix elements of $${1\over N_c}\int : q_i(x+{y\over
2})q^{\dag i}(x-{y\over 2}): e^{-ipy}dy$$ in free fermion
theory. Due to
asymptotic freedom, this is the same as the behaviour of the
interacting
theory. However in the classical theory they are to be viewed as  postulates.

We will impose
\beq
     \tilde M_d(x,p)\sim O({1\over |p|^2})\quad \tilde
M_{od}(x,p)\sim O({1\over |p|})
\eeq
where $\tilde M_d(x,p)$ (or, $\tilde M_{od}(x,p)$) is the diagonal
(or, off--diagonal) part of the $2\times 2$ matrix $\tilde M(x,p)$
in a basis where
 $\ga_5$ is diagonal. It will not be possible to absorb the central term into
the definition of $M(x,y)$ without violating these  conditions; this is what
makes the central extension non--trivial. The above commutation relations
then define a topological Lie algebra, the topology being defined by the norm
implicit in this asymptotic behaviour:
\beq
	||M||=\sup_{x,p}[(p^2+1)|\tilde M_{d}(x,p)|]+\sup_{x,p}[(|p|+1)|\tilde
M_{od}(x,p)|].
\eeq

The phase space of the theory is a co--adjoint orbit of this
Lie algebra ( or rather the corresponding Lie group, which is a  dense subgroup
of
 the
restricted Unitary group $U_1(H)$ of Segal\[pressleysegal]; we
refer to \[twodhadron] for a more detailed description). In the
case of 2dqcd, this orbit was a Grassmannian, which is defined
by a quadratic equation satisfied by $M$:
\beq
     [M+\eps_0]^2=1.
\eeq

 The  co--adjoint orbits of the restricted unitary group are
known; they are infinite dimensional analogues of the familiar
flag manifolds. ( This theory is not necessary in the following, except as
motivation for some definitions.)
 We will
require that $M$ satisfy the above constraint except for a
finite dimensional block. More precisely,that
\beq
[M+\eps_0]^2-1\;\hbox{ is finite rank}.
\eeq
 For the static solutions we are interested in,
this block is in fact one-dimensional. The meaning of this
condition is that we allow for fermion states to be
completely filled or  completely unfilled
 except for a finite dimensional block of
states which may be only partially filled. If in fact $(M+
\eps_0)^2=1$ all the
states in the theory would be either completely filled or completely unfilled
and therefore singlets under the global
 $SU(N)$ symmetry. \[geom] This  is too strong a condition in our case. In fact
the constraint that
$(M+\eps_0)^2-1$ be finite
rank will enforce that
\beq
     \tilde M^2(x,p)\sim -\sgn(p)[\ga_5,\tilde M(x,p)]_+
\eeq
or that
\beq
     \tilde M_d^2(x,p)+\tilde M_{od}^2(x,p)\sim
-\sgn(p)[\ga_5,\tilde M_{od}(x,p)]_+ .
\eeq
These asymptotic conditions will be useful later.

The hamiltonian of the theory can be obtained by  rewriting
the regularized
fermionic hamiltonian in terms of the bilinears. It is
convenient to write it in
terms of
$\tilde M(x,p)$, for then a cut--off in the range of $p$ provides a natural
regularization. After some manipulations we can bring the hamiltonian to the
form,
\beq
     E_{\Lambda}(M)=\int dx\big[\int_{|p|<\Lambda} {dp\over
2\pi} \tr
     [\ga_5 p \tilde M(x,p)+\half V(x)\tilde M(x,p)]-C\big].
\eeq
Here,
\beq
     V(x)=2 g^2(\Lambda)\int_{|p|<\Lambda} \tilde
M_{od}(x,p){dp\over 2\pi}
\eeq
and $g^2(\Lambda)$ is a coupling constant which will be picked
such that the
theory is well--defined ( see below).  Also $C$  is
picked such that
the vacuum solution has zero energy density. The equations of
motion are anyway
independent of  $C$.

It is clear from the asymptotic behaviour of $\tilde
M_{od}(x,p)$ that , if
$g^2(\Lambda)\sim {g_1^2\over \log\Lambda}$, $V(x)$ will be
independent of
$\Lambda$
 asymptotically. In fact, this will turn out to be a self--
consistent choice
for $g^2(\Lambda)$. If  moreover,
$g_1^2={\pi \over 2}$,
the quantity
\beq
    \tr [\ga_5 p\tilde M_d(x,p)+\half V(x)
\tilde M_{od}(x,p)]
\eeq
vanishes faster  than ${1\over |p|}$. ( Each term in the trace
goes like
${1\over |p|}$ but the leading contributions cancel.)
 This means that the energy can be defined by taking the limit
$\Lambda\to\infty$,
\beq
     E(M)=\int dx\big[\int {dp\over 2\pi}\tr [\ga_5 p\tilde
M(x,p)+\half V(x)
\tilde
M(x,p)]-C\big];
\eeq
a regularization is only necessary in the above expression
for $V(x)$ in terms
of $\tilde M(x,p)$.

The  equations of motion that follow from this hamiltonian and
the above
commutations relations are, ${\pdr M(x,y;t)\over \pdr
t}=\{E(M),M\}$. The Poisson bracket relations   being those of a unitary Lie
algebra,
the r.h.s. can be written in terms of the commutator of operators.
If we use operator notation, we get,
\beq
     {\pdr M(t)\over \pdr t}=[H(M),M+\eps_0]
\eeq
where the operator $H$ is defined to be the derivative of $E$
with respect to
$M$,
\beq
     H(M)={\pdr E\over \pdr M}.
\eeq
Explicitly, $H(M)$ is a differential operator,
\beq
     H(M)=-i\ga_5{\pdr \over \pdr x}+V(x)
\eeq
where $V(x)$ is related to $M$ by the equation given earlier.
Static solutions must therefore satisfy the nonlinear
equation
\beq
     [M+\eps_0,H(M)]=0.
\eeq
Since we are studying the large $N$ limit of a theory which is
exactly solvable
for every $N$, there must be  a way to solve these equations
as well: the
above
classical dynamical system must be integrable. We do not
attempt to demonstrate
the exact integrability of this system here; instead we will
obtain a family of
classical static solutions that describe a kind of topological
soliton. We will
show that the mass spectrum of this soliton agrees with the
known large $N$
limit of the mass  spectrum of the  Thirring model.

A simple solution to the above equations of motion is,
\beq
     \tilde M(x,p)+\eps_0(p)={\ga_5 p+m\ga_0\over \surd(
p^2+m^2)}
\eeq
with $V(x)=m\ga_0$. The relation between $V(x)$ and $M$ is
satisfied if
\beq
     2g^2(\Lambda)\int_{|p|<\Lambda}{dp\over 2\pi} {1\over
\surd(p^2+m^2)}=1.
\eeq
This   solution describes the vacuum, being translation
invariant. From now on we will assume that $g^2(\Lambda)$ is
given by the above equation. We have traded the dimensionless
coupling
constant $g$ for the dimensional constant $m$. The constant
$C$ is now fixed so that this solution has zero energy:
\beq
     C=2\int {dp\over 2\pi}\bigg[p\big\{{p\over \surd
(p^2+m^2)}-\sgn(p)\big\}+{m^2\over 2\surd (p^2+m^2)}\bigg]
\eeq
which is a convergent integral.

In fact the vacuum  is degenerate; we could have replaced it
by a
chirally rotated  solution,
\beq
     V(x)=e^{-i\ga_5{\theta\over 2}}m\ga_0 e^{
i\ga_5{\theta\over 2}},\quad
\quad     \tilde M(x,p)+\eps_0(p)=e^{-i\ga_5{\theta\over
2}}{\ga_5 p+m\ga_0\over
\surd( p^2+m^2)}e^{i\ga_5{\theta\over 2}}
\eeq
which would have the same energy ( zero). This suggests the
possibility of more
general solutions which  depend on $x$ and tend to two
different vacua  as
$|x|\to \pm \infty$. Such solutions would describe topological
solitons of the
bi--local theory. Notice that there is a continuous infinity
of vacua and thus
 the solitons are parametrized by an angle $\theta$,
measuring the difference
between the directions of the vacua at  infinity. We do not know
of a systematic
way to search for such static solutions of the above classical
equations for
$M$. Instead we will produce a solution by a guess inspired by
the theory of
solitons of the non--linear Schrodinger equation.

Now, $H$ is the Dirac operator of a massive spin $\half$ particle
coupled  to an
external scalar field $V(x)$. $H$ has both a  point ( discrete) spectrum
corresponding
to bound states
and a continuous spectrum corresponding to scattering states. The discrete
spectrum will have eigenvalues in the range $-m<\lambda<m$ and will have
eigenfunctions which are square integrable.The  continuous spectrum (
`scattering states') have
eigenvalues  with
$\lambda\leq -m$ or $\lambda\geq m$. There is no normalizable
eigenvector
corresponding to these; however, it is still meaningful to
speak of the
projection operator $P_I$  to the subspace with eigenvalues within
some interval $I$
of the real axis. This  spectral projection
operator  $P_I$ can be written as
\beq
     P_I=\int_I {d\lambda\over 2\pi}  \rho(\lambda)
\eeq
where the `spectral density'  $\rho(\lambda)$ is in general some operator--
valued
distribution. The contribution of a bound state will involve a
delta-function in $\lambda$ while that of the continuum will be
some continuous
function in $\lambda$. In terms of the spectral density,  we
have a decomposition,
\beq
     H=\int {d\lambda\over 2\pi} \lambda \rho(\lambda)
\eeq
which is the generalization of the familiar decomposition of a
matrix into
eigenvalues and eigenvectors. We can determine  $\rho(\lambda)$ as the
discontinuity the
resolvent
$R(\lambda)=(H-\lambda)^{-1}$ of $H$ across the real axis:
\beq
     \rho(\lambda)=\lim_{\eps\to 0^+}{1\over i}[R(\lambda+i\eps)-
R(\lambda-i\eps)].
\eeq
The resolvent in turn can be determined in terms of the Jost
solutions \[faddeev] of
scattering theory. Thus if we have an explicit expression for $V(x)$, we can in
 principle solve the scattering problem for $H$ and obtain $\rho(\lambda)$.

The static equation of motion implies that $H$ and $M+\eps_0$ commute; one way
to satisfy this is for $M+\eps_0$ and $H$ to be simultaneously diagonal.
Generically there will be no degeneracies and this will the only solution. Thus
we should expect  there to be a real valued function $\sigma(\lambda)$ such
that
\beq
	M+\eps_0=\int \sigma(\lambda)\rho(\lambda){d\lambda\over 2\pi}.
\eeq
This function must approach $\pm 1$ as $\lambda\to \pm \infty$ to satisfy the
asymptotic condition. Its value  in
a region  not contained in the spectrum of $H$ will not matter to $M$, since
$\rho(\lambda)$ will vanish there. Any discrete eigenvalue of $H$ will be
contained in an interval in which it is the only element of the spectrum. Thus
we can choose to extend $\sigma(\lambda)$ to be  a constant in a  neighborhood
of every discrete eigenvalue. This choice will turn out to be useful later.

We can regard $M(x,y)$ as describing the self--consistent way to fill the
energy states of $H$ with fermions.In fact, ${1-\sigma(\lambda)\over 2}$ is the
filling fraction, the number of states that are occupied as a fraction of the
total available ( which is $N$). For large positive $\lambda$, this must go to
zero and for large negative $\lambda$ it must go to one; there can be a finite
number of states with a fractional filling factor.

Now in our problem, $V(x)$ and $M$ are related by the self--consistency
condition,
\beq
     \int_{|p|<\Lambda} {dp\over 2\pi}\tilde M_{od}(x,p)={1\over
2g^2(\Lambda)}V(x).
\eeq
We have already fixed $g^2(\Lambda)$ from the vacuum solution, so that the
r.h.s. is logarithmically divergent. We will now show that for any $V(x)$, the
r.h.s. also contains a logarithmically divergent piece; after cancellation of
this piece, we will get a convergent equation.

First, note that by a change of variable from $p$ to $\lambda$,
\beq
	{1\over 2 g^2(\Lambda)}=\int{d\lambda \over 2\pi}{\Theta(\lambda^2-m^2)\over
k(\lambda)}\Theta(\Lambda^2+m^2-\lambda^2).
\eeq
Here $\Theta$ denotes the step function and
\beq
	k(\lambda)=\surd(\lambda^2-m^2).
\eeq
Now, the consistency condition can be written as,
\beqs{
\int {d\lambda\over 2\pi}\bigg[
\Theta(\Lambda^2-m^2){\Theta(\lambda^2-m^2)\over
k(\lambda)}V(x)-\sigma(\lambda)\int_{|p|<\Lambda}{dp\over
2\pi}\tilde
\rho_{od}(\lambda;x,p)
\bigg]=0
}\eeqs
Now,  for large $\Lambda$, the leading contribution of the second term will
come
from the region of large  $|p|$. In the limit of large $p$, the asymptotic
behaviour of
$\tilde\rho(\lambda;,p)$ will be determined by that of the resolvent symbol
$\tilde R(\lambda;x,p)$. Furthermore, the resolvent  symbol for large $|p|$ is,
\beq
	\tilde R(\lambda;x,p)\sim [\ga_5 p+V(x)-\lambda]^{-1}
\eeq
since the WKB approximation applies in this case, for smooth $V(x)$.
Thus we  find,
\beq
	\tilde \rho(\lambda;x,p)\sim \delta(\lambda-\lambda_+)\Pi_+
+\delta(\lambda-\lambda_-)\Pi_-
\eeq
 where,
\beq
	\lambda_\pm=\pm\surd(p^2+|v|^2)
\eeq
and
\beq
	\Pi_\pm={|v|^2\over |v|^2+(\lambda_\pm-p)^2}
	\pmatrix{1&{\lambda_\pm-p\over v^*}\cr
		 {\lambda_\pm-p\over v}&{(\lambda_\pm-p)^2\over |v|^2}\cr}.
\eeq
In the above, we have assumed without loss of generality that
\beq
	V(x)=\pmatrix{0&v(x)\cr v^*(x)&0\cr}
\eeq
and we have  sometimes omitted the $x,p$ dependence of $v,\lambda_\pm,\Pi_pm$
etc.  for
simplicity.  With this asymptotic expression, it can be verified that the
divergent terms in the consistency condition cancel out; so we can take the
limit as $\Lambda \to \infty$ to get
\beq
     \int {d\lambda\over
2\pi}\bigg[\sigma(\lambda)\rho_{od}(x,x;\lambda)-\Theta(\lambda^2-
m^2){V(x)\over k(\lambda)}\bigg]=0.
\eeq
We have used here, $\int {dp\over 2\pi} \tilde \rho_{od}(x,p)=\rho_{od}(x,x)$.
Although each term separately would be  divergent, the quantity in
the square brackets has a finite integral over $\lambda$.
(This can also be verified using the explicit forms given below.)

It remains now to solve the above nonlinear integral equation for $V(x)$ and
 $\sigma(\lambda)$. We will make a guess and see if it in
fact satisfies the equation.  There must be a systematic method using the
inverse scattering theory,\[faddeev] but we will not attempt to develop this
here.

We propose the ansatz,
\beq
     V(x)=m\ga_0+{m\over \eta(x)}(\ga_\theta-\ga_0)
\eeq
where,
\beq
     \eta(x)=1+e^{-\nu x},\quad \nu=2m\sin{\theta\over 2}
\eeq
and
\beq
     \ga_\theta=Q^{-1}(\theta)\ga_0 Q(\theta),\quad
Q(\theta)=e^{i\ga_5\theta/2}.
\eeq
As $x\to -\infty$, it tends to the vacuum solution, $V(x)\to m\ga_0$. As $x\to
 \infty$, it tends to another vacuum solution, differing by a chiral rotation
:$V(x)\to e^{-iga_5\theta\over 2} m\ga_0 e^{i\ga_5\theta\over 2}$. Thus this
 ansatz describes  a sort of topological soliton of our theory.

This ansatz for $V(x)$ is the well-known \[faddeev] reflectionless potential of
the Dirac
operator, which is known to be  a  soliton of the nonlinear
Schrodinger
equation. We will use these potentials to produce solitons of
our bi--local
field theory. There is no direct relationship between the
nonlinear Schrodinger
equation and our bi-local theory: indeed our theory is
relativistically
invariant  while the nonlinear Schrodinger equation is
invariant
under Galilean transformations. Yet, the static solution of both
systems involve the
same  reflectionless potential!.

The point spectrum of  $H=-i\ga_5{\pdr\over \pdr x} +V(x)$ consists of  one
normalizable eigenstate (`bound state'),
\beq
     \psi_B(x)=\surd({\nu\over 2}){e^{-\nu x\over 2}\over
\eta(x)}\pmatrix{1\cr i
e^{i{\theta\over 2}}}
\eeq
with eigenvalue $\lambda_B=m\cos {\theta\over 2}.$ Note that
$-m\leq
\lambda_B\leq m$, so that the bound state is in the `gap', the
set of values
that are forbidden as eigenvalues of the free Dirac operator. The continuum
eigenfunctions ( scattering solutions) can be written in terms of the Jost
functions,\[faddeev] and we can obtain the answer for $\rho(\lambda)$
explicitly. We omit the computations and just display the answer
\footnote{$^*$}{ from now on  we use a basis in which $\ga_5=\pmatrix{1&0\cr
 0&-1\cr},\ga_0=\pmatrix{0&-i\cr i&0\cr}$}:
\beqs{
     \rho&(x,y;\lambda)=2\pi \delta(\lambda-
\lambda_B)\psi_B(x)\psi_B^{\dag}(y)+\cr
&\Theta(\lambda^2-m^2){\sgn(\lambda)(\lambda+k(\lambda))\over 2k}
\pmatrix{\alpha(x)\alpha^*(y)+\beta^*(x)\beta(y)&
                    \beta^*(x)\alpha(y)+\al(x)\beta^*(y)\cr
     \beta(x)\alpha^*(y)+\al^*(x)\beta(y)&
                    \alpha^*(x)\alpha(y)+\beta(x)\beta^*(y)\cr
}.\cr
}\eeqs
Here,
\beq
     \al(x)=e^{ik(\lambda)x}[1+{e^{i{\phi-\theta\over 2}}-1\over
\eta(x)}],\quad
     \beta(x)={im\over
k+\lambda}e^{ik(\lambda)x}[1+{e^{i{\phi+\theta\over 2}}-1\over
\eta(x)}].
\eeq
Moreover, $\phi$ is a sort of scattering phase shift and is
determined by the
transcendental equation
\beq
     {m\over k+\lambda}\sin{\phi+\theta\over 4}=\sin{\phi-
\theta\over 4}.
\eeq

We only need  the special case $\rho_{od}(x,x)$  to verify the equation of
motion. We put in as part of the ansatz, $\sigma(\lambda)=\sgn(\lambda)$ in the
continuum, $|\lambda|\geq m$.
After some calculations the consistency condition becomes,
\beqs{
&\sigma(\lambda_B){\nu\over 2}{e^{-\nu x}\over \eta^2(x)}ie^{i{\theta\over
2}}=\cr
&\int {d\lambda\over 2\pi}\Theta(\lambda^2-m^2){im\over
k}[(1+{e^{i\theta}-1\over \eta(x)})-(1+{e^{i{\theta-\phi\over 2}}-1\over
\eta(x)})(1+{e^{i{\theta+\phi\over 2}}-1\over \eta(x)})]\cr
}\eeqs
Using ${e^{-\nu x}\over \eta^2(x)}={1\over \eta(x)}-{1\over \eta^2(x)}$  we see
that the terms are proportional to either ${1\over \eta(x)}$ or ${1\over
\eta^2(x)}$ ( Terms independent of $\eta(x)$ cancel out.)
The terms proportional to ${1\over \eta(x)}$ give, after some simplifications,
\beq
	\sigma(\lambda_B)m\sin{\theta\over 2}=-\int {d\lambda\over 2\pi} {m\over
k}\Theta(\lambda^2-m^2) 4\sin{\phi+\theta\over 4}\sin{\theta-\phi\over 4}.
\eeq
The terms proportional to ${1\over \eta^2(x)}$ happen to give the same
equation.
With $\sigma(\lambda_B)$ given by the above equation we have a solution to the
static equations of motion.

The expression for $\sigma(\lambda_B)$ can be simplified further.  If we hold
$\theta$ fixed, the scattering angle $\phi$ can be thought of as a function of
$\lambda$: either using the previous transcendental equation, or the
alternative forms,
\beq
\cot{\phi-\theta\over 4}=\cot{\theta\over 2}+{k+\lambda\over m\sin{\theta\over
4}},\quad 	\tan{\phi\over 4}={k+\lambda+m\over k+\lambda-m}\tan{\theta\over 4}.
\eeq
If we change the variable  in the integral from $\lambda$ to $\phi$, using the
differential of the first one of the above equations
$$-{ d\phi\over 4\sin^2{\phi-\theta\over 4}}=[1+{\lambda\over k}]{d\lambda\over
m\sin{\theta\over 2}},$$
we will get
\beqs{
	\sigma(\lambda_B)m\sin{\theta\over 2}&=-4\int {d\lambda\over 2\pi}[1+{\lambda
\over k}]\Theta(\lambda^2-m^2)\sin^2{\theta-\phi\over 4}\cr
&={1\over 2\pi}m{\sin\theta\over 2}
\bigg[\int_{\lambda=-\infty}^{\lambda=-m}d\phi+
\int_{\lambda=m}^{\lambda=\infty}
 d\phi\bigg].\cr
}\eeqs
Since $\phi=-\theta,0,2\pi,\theta$ respectively at
$\lambda=-\infty,-m,m,\infty$, we have
\beq
	\sigma(\lambda_B)={\theta-\pi\over \pi}.
\eeq

To summarize,  a static solution of the classical  equations of motion is
determined
by  two  real  functions $\sigma(\lambda)$ and $V(x)$ satisfying the
self--consistency  equation
\beq
     \int {d\lambda\over
2\pi}\bigg[\sigma(\lambda)\rho_{od}(x,x;\lambda)-\Theta(\lambda^2-
m^2){V(x)\over k(\lambda)}\bigg]=0
\eeq
where $\rho(\lambda)$ is  the spectral density of $H=-i\ga_5{\pdr\over \pdr
x}+V(x)$. This integral is convergent if $\sigma(\lambda)\sim \sgn(\lambda)$
for large $|\lambda|$; no cut--of is necessary. The classical static  solution
for $M$ is then
\beq
	M=-\eps_0+\int \sigma(\lambda)\rho(\lambda){d\lambda\over 2\pi}.
\eeq

We proposed the ansatz,
\beq
     V(x)=m\ga_0+{m\over \eta(x)}(\ga_\theta-\ga_0)
\eeq
where,
\beq
     \eta(x)=1+e^{-\nu x},\quad \nu=2m\sin{\theta\over 2},\quad
     \ga_\theta=Q^{-1}(\theta)\ga_0 Q(\theta),\quad
Q(\theta)=e^{i\ga_5\theta/2}.
\eeq
Then, we showed that
\beq
	\sigma(\lambda)=\sgn(\lambda)\for |\lambda|\geq m,\quad
	 \sigma(m \cos{\theta\over 2})={\theta-\pi\over \pi}
\eeq
satisfies the self--consistency relation.
(The value of $\sigma(\lambda)$ can be arbitrarily chosen for other values of
$\lambda$, since $\rho(\lambda)$ vanishes there.) Thus we have a one--parameter
family of static solutions of the classical theory. The parameter $\theta$ is a
topological  number that determines the behaviour of the solution as the center
of mass variable goes to infinity.  When $\theta=0$, we recover the vacuum
solution, $V(x)=m\ga_0$.

Although we have an explicit expression for $\rho(\lambda;x,y)$, it appears
cumbersome to evaluate the integral over $\lambda$ for $M(x,y)$. What we mean
by an exact solution is the
above integral
representation for $M$.

Next we find the energy of this configuration. The integral for energy appears
to be  too
hard to be calculated directly. We will
instead use an indirect
method. The variation of $E(M)$ under an infinitesimal change
is $\tr
H(M)\delta M$. At a static solution, $E(M)$ is invariant under
all
infinitesimal variations of $M$  satisfying the constraint but
 that do not change
its boundary conditions. However a change of $\theta$ in the
above ansatz will
change the behaviour of $M$ at infinity, and $E(M)$ need not
be stationary with
respect to this variation. ( $\theta$ is a  topological
quantum number of the
soliton.) We will get a simple expression for the derivative of energy with
respect to $\theta$.

Let us calculate   therefore
\beqs{
     {dE(\theta)\over d\theta}&=\tr H(M){d\over d\theta}(M+
\eps_0)\cr
     &=\tr H(M) \int {d\lambda\over
2\pi}[{d\sigma(\lambda)\over
d\theta}\rho(\lambda)+\sigma(\lambda){d\rho(\lambda)\over
d\theta}]\cr
}\eeqs

Now  we use the formula,
\beq
	\int {d\lambda\over 2\pi} \sigma(\lambda){d\rho(\lambda)\over d\theta}=
\int{d\lambda d\lambda'\over (2\pi)^2} {\sigma(\lambda)-\sigma(\lambda')\over
\lambda-\lambda'} \rho(\lambda){dH\over d\theta}\rho(\lambda')
\eeq
which follows from first order perturbation theory. The second term in the
expression for ${dE\over d\theta}$ becomes,
\beq
I=\int{d\lambda d\lambda'\over (2\pi)^2} {\sigma(\lambda)-\sigma(\lambda')\over
\lambda-\lambda'} \tr \rho(\lambda')H\rho(\lambda){dH\over d\theta}.
\eeq
The trace will be non--zero only if $\lambda=\lambda'$, since $H$ commutes with
$\rho(\lambda')$. Then we will have
\beq
I=\int{d\lambda \over 2\pi} {\pdr \sigma(\lambda)\over \pdr \lambda}
\tr H\rho(\lambda){dH\over d\theta}.
\eeq
Now $\sigma(\lambda)$ is a constant on the continuum, so there is no
contribution from it to this integral. As for the bound state, recall that (
for $\theta\neq 0,2\pi$) it is always separated by a gap from the continuum. We
can continue $\sigma(\lambda)$ into this gap arbitrarily since $\rho(\lambda)$
is zero there; we could  for example choose $\sigma(\lambda)$ to be constant in
some interval containing $\lambda_B$. Thus we see that $I=0$. This can probably
  also be seen
 by more direct but tedious calculations.
 We have,
\beq
   {dE(\theta)\over d\theta}=\tr H(M) \int {d\lambda\over
2\pi}{d\sigma(\lambda)\over
d\theta}\rho(\lambda)=	{1\over \pi}\lambda_B={1\over \pi}m\cos {\theta\over
2}.
\eeq
Thus we find ( recall that the configuration with $\theta=0$ is the vacuum
which has zero energy)
\beq
	E(\theta)={2\over \pi}m\sin{\theta\over 2}.
\eeq
This agrees with the large $N$ limit of the spectrum of the Thirring model as
obtained  by the Bethe ansatz method, if we identify $\mu={2\over \pi}m$.

By similar arguments we can also obtain the derivative of the energy density,
\beq
	{\pdr E(x,\theta)\over \pdr \theta}={2\over \pi}m\cos{\theta\over
2}
\psi_B^{\dag}(x)\psi_B(x)
\eeq
which leads to
\beq
	E(x,\theta)={1\over \pi}[{1\over x}\sin{\theta\over 2}\tanh(mx\sin
{\theta\over
2})-{1\over mx^2}\log(\cosh(mx\sin{\theta\over 2}))]
\eeq
This energy density is peaked around the origin and vanishes exponentially at
infinity. This is precisely what we expect for a soliton.
It should also be clear  that the soliton quantum number $\theta$ corresponds
to the
(large $N$ limit of) the discrete conserved quantum number   of the fermionic
theory.( However, the `baryon number' $B=-\tr M$ is zero for all these
configurations.
This can be verified by using the regularization methods of \[niemisem]. Thus
the situation is the opposite of that in 2dqcd; it is the abelian part of the
symmetry that is `confined'.)

There are many additional properties of these solitons that can be studied. For
example it would be interesting to obtain the time dependent solutions that
represent the  classical scattering of  several of them. It would  also be
interesting to generalize the inverse scattering  methods  of Ref. \[faddeev]
to
infinite component classical field theories such as ours. Furthermore, it
should be
possible to generalize  our results to Thirring models with several flavors.

We thank T. Turgut for useful discussions. This work was supported in part by
 by the US Department of Energy, Grant No. DE-FG02-91ER40685.

{\bf References}\hfill

\def\ni{\noindent}

\ni\gn.  D. J. Gross and A. Neveu Phys. Rev. D 10, 3235 (1974).

\ni\wiegmann. P.B. Wiegmann, Phys. Lett. 141B, 217 (1984).

\ni\andrei. N. Andrei and J.H. Lowenstein, Phys. Lett. 90B, 106 (1980)

\ni\polyakov. A. M. Polyakov,  {\it Gauge Fields and Strings} Harwood Acad.
Publ., NY(1987).

\ni\berezin. F. A. Berezin, Comm. Math. Phys. 63, 131 (1978); For a pedagogical
review see A. Perelomov, {\it Generalized Coherent States and their
Applications},
Texts and Monographs in Physics, Springer-Verlag (1986).

\ni\ohta.  M. Thies and K. Ohta, Phys. Rev. D48,5883 (1993).

\ni\twodhadron. S. G. Rajeev, in {\it 1991 Summer School in High Energy Physics
and
Cosmology}  Vol. 2  ed. E. Gava, K. Narain, S. Randjbar--Daemi, E. Sezgin and
Q. Shafi, Wold Scientific (1992); Int. J. Mod. Phys. 9, 5583  (1994).

\ni\sphqcd. K.S. Gupta, S. Guruswamy and S.G. Rajeev, Phys. Rev. D48, 3354
(1993)

\ni\others. A. Dhar, G. Mandal and S. Wadia, Phys. Lett. B239, 15 (1994);
 M. Cavicchi and P. di Vecchia, Mod. Phys.Lett. A8, 2427 (1993).

\ni\renqm. K.S. Gupta and S.G. Rajeev, Phys. Rev D48, 5940 (1993)

\ni\thooft. G. 't Hooft, Nucl. Phys. B75, 461 (1974)

\ni\twodbaryon. P. Bedaque, I. Horvath and S.G. Rajeev, Mod. Phys. Lett. A7,
3347
(1992)

\ni\araki. H. Araki, M.B. Smith and L. Smith, Comm. Math. Phys. 22, 71 (1971).

\ni\semniemi. A.J. Niemi and G.W. Semenoff, Phys. Rep. 135, 99 (1986)

\ni\fateevwieg. V. A. Fateev, V. A. Kazakov and P. B. Wiegmann, {\it  Principal
Chiral Model at Large $N$ } hepth/9403099 (1994).

\ni\kacpet. V. Kac and  D. H. Peterson, Proc. Natl. Acad. Sci. USA 78, 3308
(1981)

\ni\pressleysegal. A. Pressley and G. Segal, {\it Loop Groups}, Clarendon
Press,
Oxford (1986); J. Mickelsson, {\it Currents Algebras and Groups}, Plenum
(1989).

\ni\geom. S. G. Rajeev and T. Turgut UR preprint in preparation.

\ni\faddeev. L.D. Faddeev and L.A. Takhtajan, "Hamiltonian Methods in the
Theory of Solitons", Springer-Verlag (1980).

\bye